\documentstyle[preprint,aps,epsfig,here]{revtex}
\begin{document}

\title{Nuclear shadowing in polarized DIS on $^6$LiD at small $x$ and its effect on the extraction of the deuteron spin structure function $g_{1}^{d}(x,Q^2)$}

\author{V. Guzey \footnote{e-mail: vguzey@physics.adelaide.edu.au}}

\address{Special Research Centre for the Subatomic Structure of Matter (CSSM),\\
Adelaide University, Australia, 5005}

\preprint{
\vbox{
\hbox{ADP-00-43/T426}
}}

\maketitle

\begin{abstract}

We consider the effect of nuclear shadowing in polarized deep inelastic scattering (DIS) on $^6$LiD at small Bjorken $x$ and its relevance to  the extraction of the deuteron spin structure function $g_{1}^{d}(x,Q^2)$.
Using models, which describe nuclear shadowing in unpolarized DIS, we demonstrate that the nuclear shadowing correction to  $g_{1}^{d}(x,Q^2)$ is significant.

\end{abstract}

\section{Introduction}
\label{sec:intro}

Recent interest in the spin structure of the proton, neutron, and deuteron and advances in experimental techniques  have led to a number of experiments concerned with deep inelastic scattering (DIS) of polarized leptons on various polarized targets. Among these are the E143 experiment at SLAC \cite{E143} and the  SMC collaboration at CERN \cite{SMC}, which used polarized hydrogen and deuterium, the E154  experiment at SLAC \cite{E154} and the HERMES collaboration at DESY \cite{HERMES}, which used polarized $^3$He, and the HERMES experiment \cite{HERMESp}, which  used polarized hydrogen \cite{HERMESp}.  

A new material, deuterized lithium $^6$LiD, has recently emerged as  a source of polarized deuterium in the E155/E155x experiments at SLAC. In comparison with the previously used target materials, $^6$LiD shows a better dilution factor (polarizability) and radiation resistance (durability) \cite{li6d}.
The deuteron spin structure function $g_{1}^{d}(x,Q^2)$ was studied with the use of the polarized $^6$LiD target by the SLAC E155 experiment for the first time\cite{E155}.

In order to  extract  the spin structure functions of the proton, neutron, and deuteron from the data on polarized DIS on nuclear targets one needs to account for nuclear effects.
These effects can be divided into incoherent and 
coherent  contributions.

 The incoherent nuclear effects result from the scattering of the incoming lepton  on each individual nucleon, nucleon resonance, or  virtual meson in the nucleus. The incoherent nuclear effects  are present at all Bjorken $x$. Spin depolarization, the presence of non-nucleonic degrees of freedom, Fermi motion, binding, and off-shell effects are  examples of incoherent nuclear effects. 

In the target rest frame, coherent nuclear effects arise from the interaction of the incoming lepton with two or more nucleons of the target. The coherent nuclear effects  are typically concentrated at low values of Bjorken $x$. Nuclear shadowing at $10^{-4} \le x \le 0.05$ and subsequent  antishadowing at $0.05 \le x \le 0.2$ are  examples of coherent effects.
It is important to stress that the transition between the shadowing and antishadowing regions is not well-understood. Thus, the Bjorken $x$, where the transition occurs, is known only approximately. For example,  the NMC unpolarized DIS data on nuclei \cite{NMC} suggests that, depending on the nuclear target, the transition takes place between $x=0.02$ and $x=0.07$ \footnote{According to the NMC data \cite{NMC}, nuclear shadowing disappears at $x=0.0175$ for $^6$Li and at $x=0.07$ for $^{40}$Ca.}. Consequently, one can reliably estimate the effect of nuclear shadowing only at $x \leq 0.02$.

In  case of polarized DIS on nuclear targets, the major nuclear effect is spin depolarization. This effect manifests itself as a decrease of the effective polarization of the nucleons due to the presence of higher partial waves in bound-state nuclear wave functions \cite{BW84}.  
The effective polarization $P$ of a nucleon  is introduced as the probability 
that the nucleon carries spin of 
the fully polarized nuclear target. 
$P$ can be reliably calculated by the standard methods of nuclear physics.

As an example, one can consider the proton and neutron effective polarizations in deuteron, defined as  $P_{n}=P_{p}=1-1.5\, \omega_{D}$, where $\omega_{D}$ is the
probability of the $D$-wave in the deuteron ground-state wave function. One finds that $P_{n}=P_{p}$=0.913, using the Paris nucleon-nucleon potential \cite{Paris}, and $P_{n}=P_{p}$=0.936, using the Bonn nucleon-nucleon potential \cite{Bonn}.

As another example, one can consider the effective polarizations of the neutron and protons in $^3$He.
Calculations of the $^3$He bound-state wave function with various nucleon-nucleon potentials and three-nucleon forces yield significant probabilities of higher partial waves.  The probabilities  result in the following effective polarizations: $P_{n}=0.86 \pm 0.02$ for the neutron  and  $P_{p}=-0.028 \pm 0.004$ for each proton \cite{FGPBC}.

In case of the target of $^6$LiD, polarized deuterium originates from free deuterons D as well as from  $^6$Li since the latter can be visualized as an alpha particle plus a polarized deuteron. Treating $^6$Li as a cluster $\alpha+p+n$, the Faddeev equation for the three-body system   can be used to calculate the properties of  the ground-state wave function of $^6$Li. 
The calculations of Ref.\ \cite{Schell} indicate that the 
effective polarizations of the proton and neutron in polarized $^6$Li  are $P_{n}=P_{p}=0.866 \pm 0.012$ \cite{li6d}. In addition to this effect,
 an isotopic analysis  of the $^6$LiD target revealed that 4.6\% of lithium is $^7$Li and that 2.4\% of deuterium is hydrogen \cite{li6d}. And, finally, the effective  polarization of $^6$Li  was measured to be 97\% of the polarization of the free deuterons in $^6$LiD \cite{E155}. Thus, assuming that $^6$LiD is fully polarized with $P_{^6{\rm LiD}}=1$, the effective polarization of deuterium  in $^6$LiD is 
\begin{equation}
P_{d}=\frac{1}{2}\Big(0.976+\frac{0.97 \times 0.954 \times 0.866}{1-1.5\, \omega_{D}}\Big)=0.916 \div 0.927 \pm 0.013 \ , 
\label{pd}
\end{equation}
where the first (second) value  is for the Bonn (Paris) nucleon-nucleon potential. 
 
Also, in order to extract the precise shape of the proton, neutron, or deuteron spin structure function $g_{1}(x,Q^2)$ from the DIS data on polarized nuclear targets, one must account for the Fermi motion, binding, and off-shell effects. However, calculations for deuterium \cite{MPT95} and for $^3$He \cite{CSPS} indicate that these effects are negligible at $x \le 0.7$. Thus,  with  good accuracy one can neglect them while extracting  $g_{1}^{d}(x,Q^2)$ from the $^6$LiD data at $x \le 0.7$. 

The importance of the deuteron spin depolarization in $^6$LiD,  see Eq.\ (\ref{pd}),
 is well-established and has been  taken into account in analyzing  the data of the E155 experiment \cite{E155}.
 However, since some of the data covers the interval 
of small Bjorken $x$, $0.014 \le x \le 0.2$, corrections should be made for  nuclear shadowing and antishadowing. As explained above, this region of $x$ corresponds to the transition between the regimes of nuclear shadowing and antishadowing, which is known very poorly at the moment. Thus, one cannot estimate the effect of nuclear shadowing at $x \geq 0.02$ until a theory of antishadowing exists. Consequently, in this work, we 
estimate the effect of nuclear shadowing 
in polarized DIS on $^6$LiD
 and its influence on  the extraction of $g_{1}^{d}(x,Q^2)$
at very small Bjorken $x$ only, $10^{-4} \leq x \leq 0.02$. Our analysis is applicable only to the lowest E155 point $\langle x \rangle$=0.015.

\section{Nuclear shadowing and antishadowing effects}
\label{sec:main}

As explained in the Introduction, the polarization of $^6$LiD is formed by the effective polarizations of deuterons, $^6$Li, protons, and $^7$Li. Neglecting the Fermi motion, binding, and off-shell  effects, the spin structure function of $^6$LiD $g_{1}^{^6{\rm LiD}}(x,Q^2)$ can be written as
\begin{eqnarray}
g_{1}^{^6{\rm LiD}}(x,Q^2)&=&0.976\,g_{1}^{d}(x,Q^2)+0.97 \times 0.954\,g_{1}^{^6{\rm Li}}(x,Q^2) \nonumber\\
&+&0.024\,g_{1}^{p}(x,Q^2)+0.046 \times 0.97 \,g_{1}^{^7{\rm Li}}(x,Q^2) \ .
\label{cnvl}
\end{eqnarray}
Eq.\ (\ref{cnvl}) assumes that the admixtures of hydrogen and $^7$Li to $^6$LiD    are 100\% and 97\% polarized, respectively. 

Eq.\ (\ref{cnvl}) neglects the nuclear shadowing and antishadowing  corrections. Their importance in unpolarized and polarized DIS at small Bjorken $x$ is well-understood (for a recent review see \cite{PW99}). 
In the laboratory reference frame, nuclear shadowing arises from  the interaction of the incoming lepton with two and  more nucleons of the target. These multiple interactions 
decrease total inclusive cross sections in unpolarized DIS as well as spin asymmetries in polarized DIS. The latter fact results in a decrease of  $g_{1}^{^6{\rm LiD}}(x,Q^2)$:
\begin{eqnarray}
g_{1}^{^6{\rm LiD}}(x,Q^2)&=&0.976\,g_{1}^{d}(x,Q^2)+0.925\,g_{1}^{^6{\rm Li}}(x,Q^2)  \nonumber\\
&+&0.024\,g_{1}^{p}(x,Q^2)+0.045\,g_{1}^{^7{\rm Li}}(x,Q^2) \nonumber\\
&-&0.976\,\delta g_{1}^{d}(x,Q^2)-0.925\,\delta g_{1}^{^6{\rm Li}}(x,Q^2)-0.045\,\delta g_{1}^{^7{\rm Li}}(x,Q^2) \ , 
\label{shad}   
\end{eqnarray}
where $\delta g_{1}^{d}(x,Q^2)$, $\delta g_{1}^{^6{\rm Li}}(x,Q^2)$, and $\delta g_{1}^{^7{\rm Li}}(x,Q^2)$ denote the shadowing corrections for the corresponding spin structure functions. Thus, Eq.\ (\ref{shad}) describes the $^6$LiD spin structure function $g_{1}^{^6{\rm LiD}}(x,Q^2)$ and the corrections associated with nuclear shadowing at small Bjorken $x$, $10^{-4} \div 10^{-3} \le x \le 0.02 \div 0.05$ 
\footnote{The upper limit for nuclear shadowing $x \approx 0.05$ is usually estimated as the Bjorken $x$, when the coherence length $l_{c}=1/(2m_{N}x)$ is equal to  1.7 fm, the average internucleon distance in nuclei.}.

The amount of nuclear shadowing in Eq.\ (\ref{shad}) is expected to be significant for  two reasons. Firstly, shadowing corrections 
to the spin dependent structure functions $g_{1}(x,Q^2)$ 
are about twice as large as 
to the spin-averaged structure functions $F_{2}(x,Q^2)$
\cite{FGS96,GS99,EPW}. Secondly, shadowing corrections
are larger for heavier nuclei.
Since almost half of the polarized deuterons in $^6$LiD originate from $^6$Li,
where the nuclear shadowing correction  is significant, the shadowing corrections to  $g_{1}^{^6{\rm Li}}(x,Q^2)$ are larger than to  $g_{1}^{d}(x,Q^2)$.

The shadowing corrections for the deuteron spin structure function were calculated in Ref.\ \cite{EPW}. It was found that the ratio $\delta g_{1}^{d}(x,Q^2)/(g_{1}^{p}(x,Q^2)+g_{1}^{n}(x,Q^2))$ is, for example,   5.8\% at $x=10^{-3}$ and 4.8\% at $x=10^{-2}$.

Within the framework developed in Refs.\ \cite{FGS96,GS99}, one can estimate 
the nuclear shadowing corrections to $g_{1}^{^6{\rm Li}}(x,Q^2)$ and  $g_{1}^{^7{\rm Li}}(x,Q^2)$ using the Gribov-Glauber multiple scattering formalism along with  simple ground-state wave functions of $^6$Li and $^7$Li.
The details of this calculation are presented in appendix A.

The amount of nuclear shadowing depends on the effective cross section of the incoming photon-nucleon interaction $\sigma_{eff}$, see Eq.\ (\ref{m1}). 
We have considered two representative examples of $\sigma_{eff}$ existing in the literature. These are the $\sigma_{eff}$, which can be inferred from the two-phase model of nuclear shadowing of Ref. \cite{MT93}, and  
the $\sigma_{eff}$ from the leading-twist diffraction-based approach to nuclear shadowing of Ref. \cite{FS99}. The main difference between the two models is that the model of \cite{FS99} accounts for antishadowing by requiring that the corresponding $\sigma_{eff}$ vanishes at $x \geq 0.02$.
 Figure\ \ref{fig:one} represents $\sigma_{eff}$ of \cite{MT93} as a solid line and  $\sigma_{eff}$ of \cite{FS99} as a dashed line.

At very low Bjorken $x$,  calculations with both $\sigma_{eff}$ predict a similar amount of nuclear shadowing. Namely, at $x=10^{-4} \div 10^{-3}$ and $Q^2=4$ GeV$^2$,  the ratio  $\delta g_{1}^{^6{\rm Li}}(x,Q^2)/(g_{1}^{p}(x,Q^2)+g_{1}^{n}(x,Q^2))$ equals $0.17 \div 0.15$ ($0.12 \div 0.10$) for $\sigma_{eff}$ given by the solid (dashed) curve in Fig.\ \ref{fig:one}. 
However, at larger $x$, the deviation between the predictions made with the two scenarios for $\sigma_{eff}$ becomes larger. While, for example, at $Q^2=4$ GeV$^2$ and $x=10^{-2}$, $\delta g_{1}^{^6{\rm Li}}(x,Q^2)/(g_{1}^{p}(x,Q^2)+g_{1}^{n}(x,Q^2))=0.12$ for the calculation with the  $\sigma_{eff}$ of \cite{MT93}, $\delta g_{1}^{^6{\rm Li}}(x,Q^2)/(g_{1}^{p}(x,Q^2)+g_{1}^{n}(x,Q^2))=0.03$  for the calculation with the  $\sigma_{eff} \cite{FS99}$.

Note also that at even larger $x$, $x \approx 0.02 \div 0.05$, the calculations of nuclear shadowing bear a significant theoretical uncertainty. At those values of  $x$, the coherent length of the incident photon becomes comparable to the average internucleon distance in nuclei and, as a consequence, nuclear shadowing rapidly decreases and gives up its place to antishadowing. The position and shape of this transition is unknown. Thus, we estimate the effect of nuclear shadowing only at $10^{-4} \leq x \leq 0.02$.

  The shadowing correction to the $^7$Li  spin structure function,
given by Eq.\ (\ref{m2}), is not only sizable but also does not vanish when $g_{1}^{d}(x,Q^2)$ vanishes. Thus, at those Bjorken $x$, where $g_{1}^{d}(x,Q^2)$ is small and, hence, $\delta g_{1}^{d}(x,Q^2)$ and $\delta g_{1}^{^6{\rm Li}}(x,Q^2)$ are small, the term proportional to $\delta g_{1}^{^7{\rm Li}}(x,Q^2)$ in Eq.\ (\ref{shad}) gives the dominant contribution to the shadowing correction  to $g_{1}^{^6{\rm Li D}}(x,Q^2)$ regardless the fact that $^7$Li is only a 4.6\% admixture to $^6$LiD. At other $x$, $\delta g_{1}^{^7{\rm Li}}(x,Q^2)$ in Eq.\ (\ref{shad}) can be safely neglected.

Nuclear shadowing at $10^{-4} \le x \le 0.05$ is followed by some antishadowing at $0.05 \le x \le 0.2$, which enhances 
$g_{1}^{^6{\rm LiD}}(x,Q^2)$ above the impulse approximation prediction of Eq.\ (\ref{cnvl}). 
 In unpolarized DIS, the presence of this enhancement for the nuclear structure function $F_{2A}(x,Q^2)$ and the gluon and valence quark parton densities in nuclei
has firm experimental evidence, see Ref.\ \cite{PW99} for a review. However, since the understanding of the dynamics of nuclear antishadowing is lacking, it can only be  treated in a model-dependent way. For example, in Ref.\ \cite{anti}, the contribution of antishadowing was modelled using the baryon number and momentum sum rules for the nucleus.   

In polarized DIS, antishadowing is not  constrained by the baryon number and momentum sum rules. However, in the particular case of polarized DIS on mirror nuclei, one can use the generalization of the Bjorken sum rule \cite{FGS96}. Using this approach, the antishadowing contribution to the non-singlet nuclear spin structure function $g_{1}^{n.s.}(x,Q^2)$ of $^3$He \cite{FGS96,GS99} and $^7$Li \cite{GS99} was modelled. The contribution was found to 
be of the order of  $14 \div 40$\% for the $A=3$ system and of the  order of  $20 \div 55$\% for the $A=7$ system. The spread of the presented values is an indication of the uncertainty of where the transition between the shadowing and antishadowing regions takes place. 

Although the generalization of the Bjorken sum rule for $^6$LiD does not exist because $^6$LiD is an isoscalar, 
there is no reason for the absence of the antishadowing correction  in DIS  on polarized $^6$LiD.
While nuclear antishadowing is expected to modify the extraction of $g_{1}^{d}(x,Q^2)$ from the $^6$LiD data at $0.02 \div 0.05 \le x \le 0.2$,
we do not estimate this effect and simply confine our predictions to the nuclear shadowing range of $x$, $10^{-4} \le x \le 0.02$.

In this work, the shadowing correction to $g_{1}^{^6{\rm LiD}}(x,Q^2)$ is calculated using Eq.\ (\ref{shad}) at a fixed low scale $Q^2=Q_{0}^2=4$ GeV$^2$. 
In order to find the modification of $g_{1}^{^6{\rm LiD}}(x,Q^2)$ due to  the nuclear effects at larger $Q^2$, $Q^2 > Q_{0}^2$, the QCD evolution with the input, described by Eq.\ (\ref{shad}), should be used. 
Based on the  experience from the QCD evolution of unpolarized parton densities, it is expected that nuclear shadowing at small Bjorken $x$ and 
high $Q^2$ will decrease because of the contribution of the polarized gluons originating from the unshadowed, high $x$, region at the initial evolution scale $Q_{0}^2$.

Now we are in position to give an estimate of the importance of the shadowing correction in Eq.\ (\ref{shad}) on the extraction of the deuteron spin structure function  $g_{1}^{d}(x,Q^2)$.
Let us denote by $g_{1exp.}^{d}(x,Q^2)$ the deuteron spin structure function in the impulse approximation, i.e. 
$g_{1exp.}^{d}(x,Q^2)$ is 
 obtained from Eq.\ (\ref{cnvl}) where the coherent effects at small Bjorken $x$ are neglected.
  The ratio of the theoretical prediction for $g_{1}^d(x,Q^2)$, given by Eq.\ (\ref{shad}), when the effect of shadowing is present, and $g_{1exp.}^d(x,Q^2)$
is presented as
\begin{eqnarray}
&&\frac{g_{1}^d(x,Q^2)}{g_{1exp.}^d(x,Q^2)}= \nonumber\\
&&1+\frac{1}{2P_{d}\,g_{1exp.}^d(x,Q^2)}\Big(0.976\,\delta g_{1}^{d}(x,Q^2)+0.925\,\delta g_{1}^{^6{\rm Li}}(x,Q^2)+0.045\,\delta g_{1}^{^7{\rm Li}}(x,Q^2)\Big) \ ,
\label{master}
\end{eqnarray}
where $P_{d}$ is given by Eq.\ (\ref{pd}). Note that the ratio $g_{1}^d(x,Q^2) / g_{1exp.}^d(x,Q^2)$  is equal to  unity if the effect of nuclear shadowing is neglected.

It is important to stress  that the quantity, which is measured in polarized DIS, is the spin asymmetry $A_{\parallel}$, where $A_{\parallel} \approx g_{1exp.}(x,Q^2)/F_{1}(x,Q^2)$. Then, in polarized DIS, one obtains $g_{1exp.}(x,Q^2)$ by multiplying $A_{\parallel}$ by the spin-averaged structure function $F_{1}(x,Q^2)$, which is also experimentally measured and, therefore, contains all nuclear effects.  The latter fact is true for the E155 experiment data analysis \cite{Str00}, regardless the fact that  it is  
not apparent from the E155 publication \cite{E155}.  Thus, Eq.\ (\ref{master}) indeed describes the shadowing correction to the experimentally measured 
$g_{1exp.}(x,Q^2)$.

Using the results of Ref.\ \cite{EPW} for $\delta g_{1}^{d}(x,Q^2)$ and of appendix A  for $\delta g_{1}^{^6{\rm Li}}(x,Q^2)$ and $\delta g_{1}^{^7{\rm Li}}(x,Q^2)$, the ratio $g_{1}^d(x,Q^2) / g_{1exp.}^d(x,Q^2)$ of Eq.\ (\ref{master}) is presented as a function of $x$ at $Q_{0}^2$=4 GeV$^2$ in Figure\ \ref{fig:two}. The solid line is a result of the calculation  without the $\delta g_{1}^{^7{\rm Li}}(x,Q^2)$ term.
The ratio $g_{1}^d(x,Q^2) / g_{1exp.}^d(x,Q^2)$ with the $\delta g_{1}^{^7{\rm Li}}(x,Q^2)$ term included is presented as the dotted line. These two curves are obtained using $\sigma_{eff}$ of the two-phase model of Ref.\ \cite{MT93}.
The shadowing correction to $g_{1}^{^7{\rm Li}}(x,Q^2)$ was calculated by Eq.\ (\ref{m2}). Since the deuteron spin structure function parameterization of Ref.\ \cite{E155} covers the region of $x \ge 0.01$,
the contribution of the $\delta g_{1}^{^7{\rm Li}}(x,Q^2)$ term starts at $x=0.01$ in Fig.\ \ref{fig:two}.

The calculation with the leading-twist $\sigma_{eff}$ of Ref.\ \cite{FS99}
is presented by the dashed line in Fig.\ \ref{fig:two}. In this case, the term proportional to $\delta g_{1}^{^7{\rm Li}}(x,Q^2)$ does not contribute to the net shadowing correction to $g_{1}^{^6{\rm Li D}}(x,Q^2)$ because $\sigma_{eff}$ is negligibly small at $x \ge 0.01$, where $g_{1}^{d}(x,Q^2)$ is 
parameterized and sizable.

The results of the calculations with the Paris and Bonn nucleon-nucleon potentials  are virtually the same. In Fig.\ \ref{fig:two}, the Paris nucleon-nucleon potential for $\omega_{D}$ is used.

Fig.\ \ref{fig:two} illustrates that, at small Bjorken $x$, $10^{-4} \le x \le 10^{-3}$, the shadowing correction is a slow function of $x$, i.e. shadowing is saturated, and it works to increase $g_{1exp.}^{d}(x,Q^2)$ by  $13.5 \div 12$\% for the calculation with $\sigma_{eff}$ of \cite{MT93} and by $11.5 \div 10$\% for the calculation with 
$\sigma_{eff}$ of \cite{FS99}.
 At $x \ge 10^{-3}$ the shadowing correction begins to decrease more
rapidly as a function of $x$. For the lowest data point of the E155 experiment $\langle x \rangle$=0.015, 
we predict that $g_{1}^d(x,Q^2) / g_{1exp.}^d(x,Q^2)$ could still be as large as $9$\% for the calculation with $\sigma_{eff}$ of \cite{MT93}.

Thus, we conclude that nuclear shadowing does modify the extraction of the deuteron spin structure function $g_{1}^{d}(x,Q^2)$ from $g_{1}^{^6{\rm Li D}}(x,Q^2)$ at $x=10^{-4} \leq x \leq 0.02$.

  Note also that, since the E155 data \cite{E155} indicates that $|g_{1}^{d}(x,Q^2)|$ is non-zero and quite significant at small Bjorken $x$ , the shadowing correction $\delta g_{1}^{d}(x,Q^2)$  is important  for the extraction of the neutron spin structure function $g_{1}^{n}(x,Q^2)$ from $g_{1}^{d}(x,Q^2)$.

While  the present day data on $g_{1}^{d}(x,Q^2)$
is not accurate enough to  demonstrate 
the importance of nuclear shadowing, 
in the future, when  high precision data at even lower $x$ becomes available, the importance of nuclear effects typical for low Bjorken $x$  can be unambiguously established. Moreover, with high precision polarized DIS data one can study  the role played by polarized gluons. Since, in unpolarized DIS, nuclear shadowing in the gluon channel is expected to be 3 times larger than in the sea quark channel (see Fig.\ \ref{fig:one} for the the corresponding $\sigma_{eff}$), the shadowing correction to the polarized nuclear gluon parton density could be 3  times as large as the shadowing correction to the structure function $g_{1}(x,Q^2)$ \cite{GS99}.

\section{Conclusions}

The recent SLAC E155 experiment has used deuterized lithium $^6$LiD as a source of polarized deuterons in order to study the deuteron spin structure function $g_{1}^{d}(x,Q^2)$ at $0.014 \le x \le 0.9$. Since some of the data covers the region of small Bjorken $x$, $0.014 \le x \le 0.05$, where  nuclear shadowing and antishadowing play an important role, necessary corrections should be made.

In this work we considered nuclear shadowing in polarized DIS on the  $^6$LiD target and its effect on the spin structure function $g_{1}^{^6{\rm LiD}}(x,Q^2)$ 
 at small Bjorken $x$, $10^{-4} \leq x \leq 0.02$. The previous analysis of polarized DIS on deuterium, $^3$He, and $^7$Li suggests that the effect of nuclear shadowing in the nuclear spin dependent structure functions $g_{1}^{A}(x,Q^2)$ is enhanced by a factor of two as compared to the spin averaged structure functions $F_{2A}(x,Q^2)$. 

The magnitude of the  nuclear shadowing effect is represented using the ratio $g_{1}^{d}(x,Q^2)/g_{1exp.}^{d}(x,Q^2)$, see Eq.\ (\ref{master}). While, in the absence of the shadowing corrections, the ratio $g_{1}^{d}(x,Q^2)/g_{1exp.}^{d}(x,Q^2)$ equals  unity, nuclear shadowing at $x=10^{-4} \div 10^{-3}$ increases the ratio above unity   by 
 $13.5 \div 12$\% for the calculation  with the $\sigma_{eff}$ extracted from Ref.\ \cite{MT93} and by $11.5 \div 10$\% for the calculation with $\sigma_{eff}$ of \cite{FS99}.
For the lowest data point of the E155 experiment $\langle x \rangle $=0.015, 
the shadowing correction to the ratio $g_{1}^{d}(x,Q^2)/g_{1exp.}^{d}(x,Q^2)$ could be as large as $9$\%. Therefore, nuclear shadowing does modify the extraction of $g_{1}^{d}(x,Q^2)$ from $g_{1}^{^6{\rm LiD}}(x,Q^2)$ in the range of $10^{-4} \leq x \leq 0.02$ and, consequently, affects the extraction of $g_{1}^{n}(x,Q^2)$ from $g_{1}^{d}(x,Q^2)$.

Further theoretical effort is required to understand the dynamical mechanism of antishadowing, which was not discussed in this work.

Finally, we would like to stress  that  
the phenomenon of nuclear shadowing in polarized DIS on nuclear targets is a genuine low $x$ nuclear effect, which should be treated on the equal footing with any other nuclear effect, such as, for instance, spin depolarization.

\section{Acknowledgements}

I would like to thank Mark Strikman and Anthony W. Thomas for helpful comments and discussions
and Kazuo Tsushima for pointing and discussing Refs.\ \cite{Tanihata85,ADNDT}. This work was partially supported by the Australian Research Council.

\appendix
\section{Derivation of the nuclear shadowing correction for polarized DIS on $^6$Li and $^7$Li}

In this appendix, we outline key steps in the derivation of the shadowing contribution in polarized DIS on the $^6$Li and $^7$Li targets. In our analysis, we will closely follow the approach presented in Refs.\ \cite{FGS96,GS99}. 

In the laboratory reference frame, the incoming polarized photon with the high energy $\nu$, momentum $q$, four-momentum $Q^2$ and small Bjorken $x$ interacts with the hadronic target by means of its coherent quark-gluon fluctuations $|h_{i}\rangle$
\begin{equation}
 \sigma_{\gamma^{\ast} A}(\nu,Q^2)=\sum_{i}|\langle \gamma^{\ast}| h_{i} \rangle|^2 \sigma_{h_{i} A}(\nu,Q^2) \ ,
\label{eq1}
\end{equation}
where $\sigma_{\gamma^{\ast} A}$ and $\sigma_{h_{i} A}$ are the photon- and $| h_{i} \rangle$-nucleus cross sections, respectively; $|\langle \gamma^{\ast}| h_{i} \rangle|^2$ is the probability to find the configuration $| h_{i} \rangle$ in the photon wave function.

Following Refs.\ \cite{FGS96,GS99}, we have replaced the sum in Eq.\ (\ref{eq1}) by an effective fluctuation $|h_{eff} \rangle$ with $M_{h_{eff}}^2 \approx Q^2$ and the $|h_{eff} \rangle$-nucleon scattering cross section $\sigma_{eff}$. 
We have also made a hypothesis that $\sigma_{eff}$ in polarized DIS is the same as in the unpolarized DIS. We considered two models for $\sigma_{eff}$.

\begin{figure}
\hspace{-2cm}
\includegraphics[scale=.9]{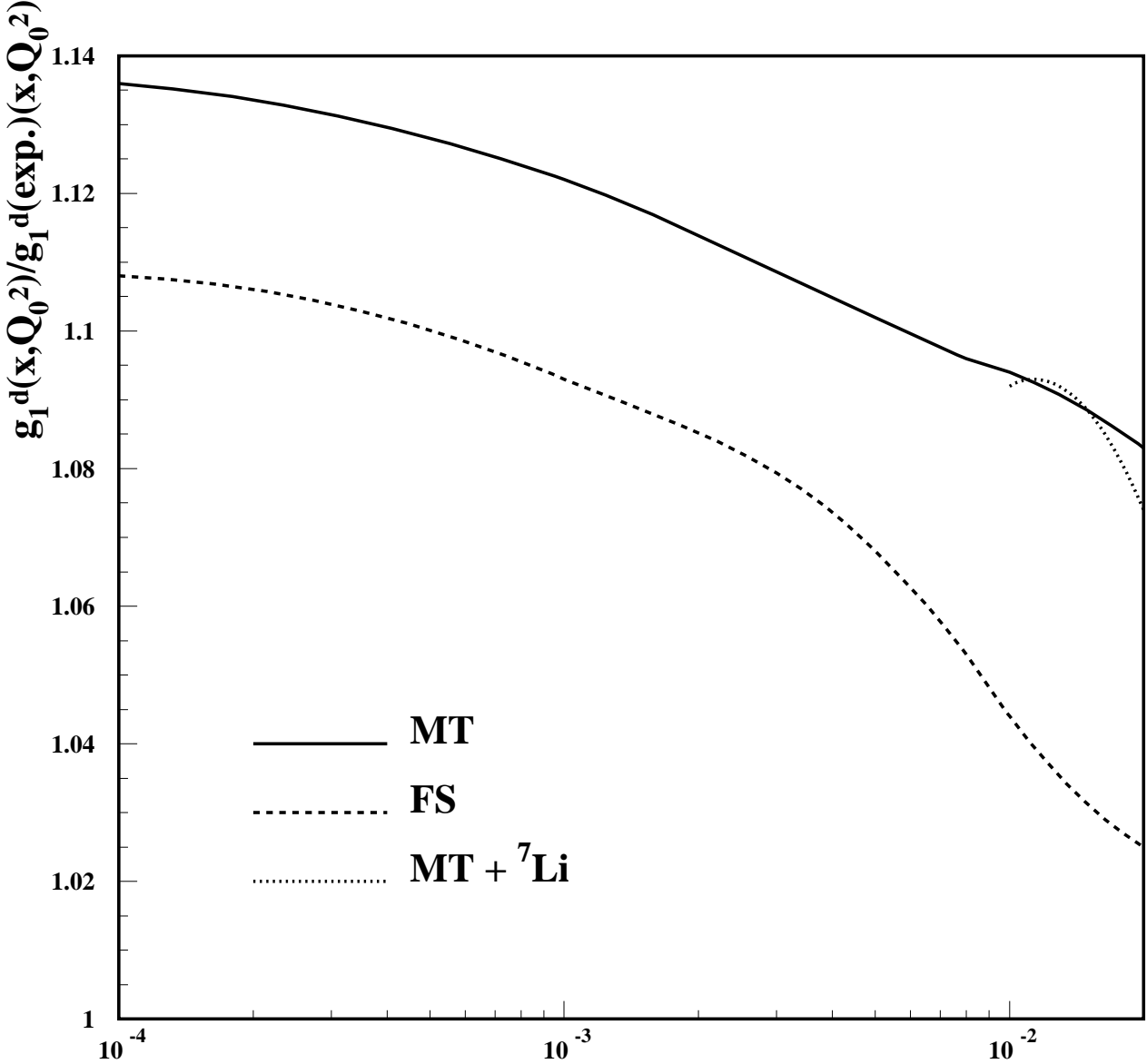}
\caption{The ratio $g_{1}^{d}(x,Q^2)/g_{1exp.}^{d}(x,Q^2)$ of Eq.\ (\ref{master})  as a function of $x$ at $Q_{0}^2$=4 GeV$^2$. The solid line is the result of the calculation with $\sigma_{eff}$ of Ref.\ [21] without the $\delta g_{1}^{^7{\rm Li}}(x,Q^2)$ term in Eq.\ (\ref{master}). $g_{1}^{d}(x,Q^2)/g_{1exp.}^{d}(x,Q^2)$ with  the $\delta g_{1}^{^7{\rm Li}}(x,Q^2)$ term included, where applicable, is given by the dotted line.
The calculation with $\sigma_{eff}$ of Ref.\ [22]
is presented as a dashed line. 
}
\label{fig:two}
\end{figure}

Using the connection between the leading contribution to nuclear shadowing, proportional to $\sigma_{eff}$, and   diffractive scattering on the proton, $\gamma^{\ast}+p \rightarrow X +p^{\prime}$, the leading-twist model for  $\sigma_{eff}$ was derived in Ref.\ \cite{FS99}. The  dashed line, denoted as ``FS'', presents the corresponding $\sigma_{eff}$ 
as a function of $x$ at $Q_0$=2 GeV in  Fig.\ \ref{fig:two}. 
 
Note also that  a similar value of $\sigma_{eff}$ at $x=10^{-3}$ and $Q^2$ equal a few GeV$^2$ can be extracted from the analysis of nuclear shadowing in unpolarized DIS on nuclei with $A \ge 12$.

The two-phase  model of nuclear shadowing of Ref.\ \cite{MT93} was successfully applied in unpolarized DIS on light and heavy nuclei in order  to describe the experimental data on the ratio $F_{2A}/F_{2D}$. The corresponding $\sigma_{eff}$ contains both the leading-twist (Pomeron and triple Pomeron) and sub-leading twist (vector mesons) contributions. It is presented as a solid line in  Fig.\ \ref{fig:two} and is denoted as ``MT''. 

Thus, we used the models for $\sigma_{eff}$ of Refs.\ \cite{FS99} and \cite{MT93} as  estimates of the lower and upper limits  for the amount of nuclear shadowing in polarized DIS on $^6$Li and $^7$Li.

 Within the discussed  approximation, the photon-nucleus cross section is proportional to the $|h_{eff} \rangle$-nucleus cross section:
 \begin{equation}
\sigma_{\gamma^{\ast} A}(\nu,Q^2) \propto \sigma_{h_{eff} A}(\nu,Q^2) \ ,
\label{eq4}
\end{equation}

The latter can be calculated using the Gribov-Glauber multiple scattering formalism \cite{Glauber,Gribov}, generalized to include the non-zero longitudinal momentum transferred to the target $q_{\parallel} \approx 2 m_{N} x$, which leads to an $x$ dependence of nuclear shadowing.  
The Gribov-Glauber  scattering formalism requires the  knowledge of the nuclear ground-state wave function and the elementary $|h_{eff} \rangle$-nucleon scattering amplitude.

 Simple forms for the ground-state wave functions of polarized $^6$Li and $^7$Li are assumed. In $^6$Li, 
one proton and one neutron are polarized with the effective polarization $P=0.866$. The effective polarization of the valence nucleons (predominantly the proton) of the $^7$Li ground-state wave function is given by the nuclear shell-model expression \cite{GS99}. In addition, the distribution of the nucleons in the configuration space, given by the square of the corresponding wave function, is taken as a simple Gaussian shape 
\begin{equation}
|\Psi_{^6{\rm Li},^7{\rm Li}}|^2 \propto \exp(-\frac{3}{2} \frac{r^2}{\langle r^2 \rangle}) \ ,
\label{wf}
\end{equation}
where $\langle r^2 \rangle$ is the average electromagnetic radius of the nucleus. $\langle r^2 \rangle ^{1/2}=2.56 \pm 0.10$ fm for $^6$Li \cite{Tanihata85} and $\langle r^2 \rangle ^{1/2}=2.41 \pm 0.10$ fm for $^7$Li \cite{ADNDT}. Note that nucleon-nucleon correlations in the nuclear wave functions (\ref{wf}) are neglected. This is a good approximation for the low Bjorken $x$, $10^{-4} \leq x \leq 0.02$, considered in this work.

The $|h_{eff} \rangle$-nucleon scattering amplitude is chosen to be purely imaginary with $B$=6 GeV$^{-2}$ being the slope of the $|h_{eff} \rangle$-nucleon cross section.

Keeping the double and triple scattering contributions, the nuclear shadowing correction to the spin structure function $g_{1}^{^6{\rm Li}}(x,Q^2)$ can be presented as
\begin{eqnarray}
&&\delta g_{1}^{^6{\rm Li}}(x,Q^2)=\Big(5\frac{\sigma_{eff}(x)}{4\pi(\langle r^2 \rangle/3+B)} e^{-q_{\parallel}^2 \langle r^2 \rangle /3}-10 \frac{\sigma_{eff}(x)^2}{48\pi^2(\langle r^2 \rangle/3+B)^2}g(x)\Big) \nonumber\\
&&\times P_{n} \Big(g_{1}^{p}(x,Q^2)+g_{1}^{n}(x,Q^2)\Big) \ ,
\label{m1}
\end{eqnarray} 
where $g(x)$ is a weak function of $x$, normalized as $g(0)$=1.

The shadowing correction, described by Eq.\ (\ref{m1}), depends on the value of $\sigma_{eff}$. Using the ``MT'' (``FS'') scenario for $\sigma_{eff}$, one finds  that the ratio   $\delta g_{1}^{^6{\rm Li}}(x,Q^2)/(g_{1}^{p}(x,Q^2)+g_{1}^{n}(x,Q^2))$ equals 0.17 (0.12) at $x=10^{-4}$, 0.15 (0.10) at $x=10^{-3}$, and 0.12 (0.3) at $x=0.01$. 

Using Eq.\ (7) of Ref.\ \cite{GS99} along with  the numerical values for the corresponding constants,  the nuclear shadowing correction to the spin dependent structure function of $^7$Li, $\delta g_{1}^{^7{\rm Li}}(x,Q^2)$, can be presented in the following form
\begin{eqnarray}
&&\delta g_{1}^{^7{\rm Li}}(x,Q^2)=\sigma_{eff}\Bigg(7.62 \times 10^{-3}\,g_{1}^{p}(x,Q^2)+1.51\times 10^{-3}\Big(g_{1}^{p}(x,Q^2)+g_{1}^{n}(x,Q^2)\Big)\Bigg)e^{-176\,x^2} \nonumber\\
&&-(\sigma_{eff})^2\Bigg(4.08 \times 10^{-5}\,g_{1}^{p}(x,Q^2)+8.65\times 10^{-6}\Big(g_{1}^{p}(x,Q^2)+g_{1}^{n}(x,Q^2)\Big)\Bigg)g(x) \ ,
\label{m2}
\end{eqnarray}
where $g(x)$ is a slow function of $x$, normalized as $g(0)=1$. Eq.\ (\ref{m2}) includes the double and triple scattering contributions to $\delta g_{1}^{^7{\rm Li}}(x,Q^2)$. Higher multiple scattering order terms are negligibly small. 

\begin{figure}
\hspace{-2cm}
\includegraphics[scale=.9]{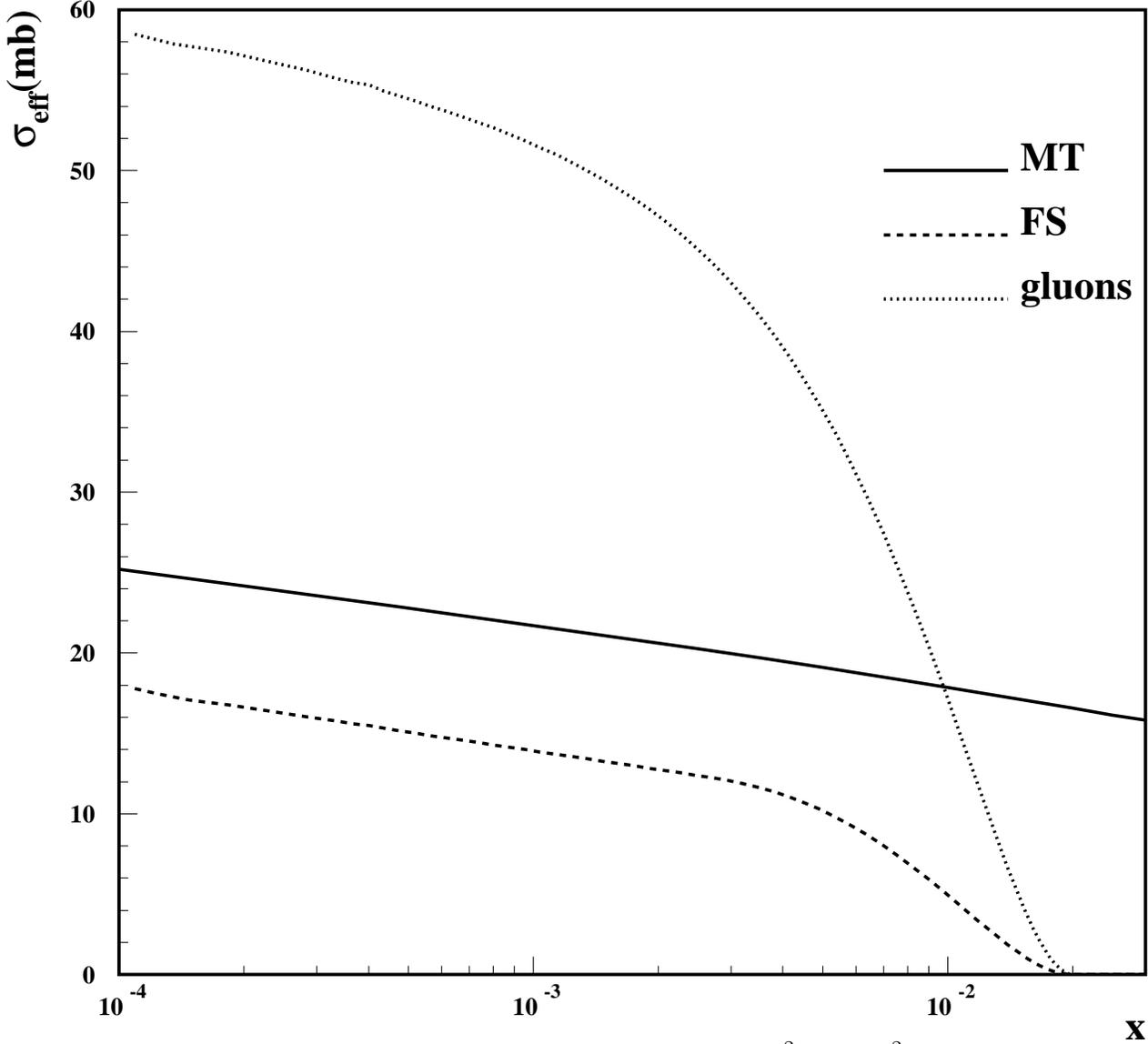}
\caption{Two scenarios for $\sigma_{eff}$ as a function of $x$ at $Q_{0}^2$=4 GeV$^2$. The solid line represents $\sigma_{eff}$ inferred 
from the two-phase model of Ref.\ [21]. The dashed line is from 
 the leading-twist diffraction-based picture of nuclear shadowing  of Ref.\ [22].  The dotted line represents $\sigma_{eff}$ of [22] for the gluon-induced nuclear shadowing.}
\label{fig:one}
\end{figure}

Unlike $\delta g_{1}^{d}(x,Q^2)$ and  $\delta g_{1}^{^6{\rm Li}}(x,Q^2)$, $\delta g_{1}^{^7{\rm Li}}(x,Q^2)$ is not proportional to $g_{1}^{p}(x,Q^2)+g_{1}^{n}(x,Q^2)$. Thus, at those Bjorken $x$, where $g_{1}^{p}(x,Q^2)+g_{1}^{n}(x,Q^2)$ and, hence, $\delta g_{1}^{d}(x,Q^2)$ and  $\delta g_{1}^{^6{\rm Li}}(x,Q^2)$ vanish, in spite of $^7$Li being a small admixture to $^6$LiD, $\delta g_{1}^{^7{\rm Li}}(x,Q^2)$ becomes significant (see the dotted line in Fig.\ \ref{fig:one}).

\end{document}